\documentclass{article}

\usepackage{PRIMEarxiv}

\usepackage[utf8]{inputenc} 
\usepackage[T1]{fontenc}    
\usepackage{hyperref}       
\usepackage{url}            
\usepackage{booktabs}       
\usepackage{amsfonts}       
\usepackage{nicefrac}       
\usepackage{microtype}      
\usepackage{lipsum}
\usepackage{fancyhdr}       
\usepackage{graphicx}       
\graphicspath{{media/}}     
\usepackage{gensymb}        
\usepackage{amsmath}

\title{Generative diffusion-based downscaling for climate}
\author{
  R. A. Watt\thanks{Equal contribution} , L. A. Mansfield\footnotemark[1]\\
  Stanford University \\
  Palo Alto, California \\
  \texttt{\{rwatt1, lauraman\}@stanford.edu}
}

\begin{document}
\maketitle

\begin{abstract}
Downscaling, or super-resolution, provides decision-makers with detailed, high-resolution information about the potential risks and impacts of climate change, based on climate model output. Machine learning algorithms are proving themselves to be efficient and accurate approaches to downscaling. Here, we show how a generative, diffusion-based approach to downscaling gives accurate downscaled results. We focus on an idealised setting where we recover ERA5 at $0.25\degree$~resolution from coarse grained version at $2\degree$~resolution. The diffusion-based method provides superior accuracy compared to a standard U-Net, particularly at the fine scales, as highlighted by a spectral decomposition. Additionally, the generative approach provides users with a probability distribution which can be used for risk assessment. This research highlights the potential of diffusion-based downscaling techniques in providing reliable and detailed climate predictions.

\end{abstract}


\section{Introduction}

Climate change poses a threat to humans and ecosystems all over the world. The potential impacts range from increased risk of damage caused by extreme events such as heatwaves, heavy rainfall, and flooding events, to disruptions in biodiversity, agriculture, and threats to food security. This underscores the critical need for accurate predictions of future climate conditions for effective mitigation and adaptation strategies. 

Global Climate Models (GCMs) have been pivotal in our efforts to simulate the Earth's complex system and project future climate scenarios. GCMs are sophisticated computer models that incorporate components from the atmosphere, ocean, land, vegetation, and sea-ice, creating a coupled Earth system. However, the computational cost associated with running GCMs at high resolutions presents a significant challenge. Currently, GCMs operate on a grid with a typical resolution of approximately $1\degree$ (roughly 100 kilometers) \cite{chen_framing_2021}. For assessing climate change impacts, we often require more localised predictions, crucial for addressing regional impacts. This limitation prompts the exploration of alternative methodologies, such as downscaling, to enhance the spatial precision of climate predictions.

In recent years, machine learning techniques have proven capable of making substantial strides in predicting climate patterns. Many believe it will revolutionise our approach to climate modeling, by offering computationally efficient methods to improve, or even replace, traditional GCMs \cite{balaji_climbing_2021,schneider_harnessing_2023,mansfield_updates_2023,rolnick_tackling_2023}. The challenges of climate and weather prediction have caught the attention of large technology companies, such as Google, NVIDIA, Microsoft, Huawei, and others, who are leveraging machine learning to facilitate numerical weather prediction \cite{lam_learning_2023,pathak_fourcastnet_2022,bi_pangu-weather_2022,nguyen_climax_2023,chen_fuxi_2023}. There is also a growing interest in machine learning emulators of GCMs \cite{mansfield_predicting_2020,watson_machine_2022,watt-meyer_ace_2023} and in embedding machine learning components into GCMs (such as subgrid-scale parameterisations) \cite{kochkov_neural_2024,ogorman_using_2018,gentine_could_2018}, amongst other uses \cite{molina_review_2023}.

One specific avenue gaining popularity is downscaling, a technique where coarse-resolution climate models are refined using machine learning to generate high-resolution predictions. This could enhance the cost-effectiveness of coarse climate models and their precision on regional scales. This paper uses a generative machine learning approach based on diffusion \cite{sohl-dickstein_deep_2015} for downscaling climate data. Using the ERA5 reanalysis dataset, we show how a diffusion models can enhance a coarse resolution map of temperature and winds over the USA \cite{hersbach_era5_2020}. The benefit is that ensembles can be produced, which are crucial for assessing model uncertainty in climate change studies \cite{hawkins_potential_2009}. Diffusion-based generative models have been applied in similar downscaling applications \cite{mardani_generative_2023,addison_machine_2022,nath_forecasting_2024}, although they have not yet gained prominence in the climate downscaling literature. Our aim is to contribute to the evolving landscape of machine learning climate downscaling, with a diffusion-based approach applied on continental scales for one of the first times. While previous studies focus on downscaling to km-scale resolution \cite{mardani_generative_2023,addison_machine_2022,nath_forecasting_2024}, we envision how diffusion could be used to downscale conventional climate models with typical CMIP-resolution, i.e., $\mathcal{O}($1\degree$)$ or $\mathcal{O}(100\mbox{ km})$, to a higher resolution that would typically require a regional climate model, i.e.,$\mathcal{O}($0.1\degree$)$ or $\mathcal{O}(10\mbox{ km})$. In agreement with several recent studies, our results show that diffusion-based generative models are a promising approach for climate data, both for downscaling 
\cite{mardani_generative_2023,addison_machine_2022,nath_forecasting_2024} and in other applications \cite{huang_diffda_2024,price_gencast_2023,chan_hyper-diffusion_2024,bassetti_diffesm_2023,cachay_dyffusion_2023}.

\section{Background}
\label{sec:background}

\subsection{Downscaling}
\label{sec:Downscale}
Climate downscaling, often referred to as super-resolution (based on the machine vision literature), is the process of refining predictions from a low-resolution climate model to a higher resolution. This step is crucial for addressing the limitations of global climate models (GCMs) and tailoring climate predictions to local or regional scales. Traditionally, two main approaches have been employed for downscaling: dynamical downscaling and statistical downscaling, where the latter includes machine learning techniques growing in popularity recently.

\textbf{Dynamical downscaling} involves running a regional climate model (RCM) at a higher resolution, typically 10-50 km, over a specific region of interest \cite{sunyer_comparison_2012,giorgi_thirty_2019,tapiador_regional_2020,giorgi_regional_2015}. The low-resolution output from a GCM serves as both the boundary and initial conditions for the RCM. RCMs offer the advantage of enhanced spatial resolution, while guaranteeing the output is dynamically consistent, rendering them popular amongst stakeholders and policymakers for important decisions \cite{gutowski_ongoing_2020}. However, RCMs also have inherent biases, like GCMs, and may still require postprocessing techniques to remove biases. Importantly, dynamical downscaling comes with a notable drawback in the high computational cost of running high-resolution RCMs.

\textbf{Statistical downscaling}, in contrast, utilises statistical methods to establish relationships between coarse-resolution model outputs and high-resolution observations in historical data \cite{maraun_statistical_2018,vandal_intercomparison_2019}. One approach is to employ a regression model which directly predicts high resolution model variables from low resolution model variables. This assumes that the low resolution model is a ``perfect'' predictor of the high resolution model (i.e., there are no model biases). This is traditionally known as ``perfect prognosis'' \cite{schubert_downscaling_1998}. Since low resolution models are often missing processes and feedbacks, this approach alone can lead to inaccurate downscaled predictions \cite{maraun_perfect_2018}. An alternative approach, known as ``model output statistics'', aims to match the predicted statistics to observed statistics, such as the mean, standard deviation, and/or quantiles. This automatically accounts for model biases and is used for the NASA Earth Exchange Global Daily Downscaled Projections \cite{thrasher_nasa_2022}. Another approach lies in ``stochastic weather generators'', that simulate stochasticity of weather data based on characteristics of observations \cite{wilby_statistical_1998} 

Although faster than dynamical downscaling, these traditional methods often exhibit poor performance in extrapolating to new climates. This limitation arises from the assumption that the same statistical relationships hold under changing conditions, known as stationarity, which we cannot necessarily expect to be true \cite{fowler_linking_2007,maraun_towards_2017}. One solution could be hybrid downscaling, such as statistical downscaling applied to RCM outputs \cite{sunyer_comparison_2012}. Alternatively, machine learning/artificial intelligence may offer solutions to this problem.


\textbf{Machine learning} techniques are becoming a popular choice for statistical downscaling that can be framed as perfect prognosis \cite{bano-medina_downscaling_2022}, model output statistics \cite{pour_model_2018}, weather generators \cite{wilby_statistical_1998} or a combination of these \cite{vandal_intercomparison_2019}. Many of these approaches are based on developments in computer vision for super-resolution of images \cite{glasner_super-resolution_2009}.


\subsection{Machine learning}
\label{sec:ML-downscale}

Downscaling can be viewed as a supervised learning task, where the goal is to learn the variables on the high resolution grid (the target), from the variables on the low resolution  grid (the input). The majority of machine learning studies aim to directly learn the mapping between the low resolution input and the high resolution target with a wide variety of machine learning architectures, including random forests \cite{hasan_karaman_evaluation_2023,medrano_downscaling_2023,chen_easy--use_2021}, support vector machines \cite{pour_model_2018}, and convolutional neural network (CNN) architectures \cite{bano-medina_downscaling_2022,jiang_downscaling_2021,wang_end--end_2019}, including U-Net architectures \cite{agrawal_machine_2019,kaparakis_wf-unet_2023,adewoyin_tru-net_2021,cho_new_2023}, convolutional autoencoders \cite{babaousmail_novel_2021} and fourier neural operators \cite{yang_fourier_2023}. These studies show that machine learning can be more accurate than conventional statistical downscaling, with the benefit of low computational cost compared against dynamical downscaling approaches. However, end-users are often concerned when there is a risk of unrealistic predictions from black-box machine learning methods, especially when used in new scenarios such as climate change. Recently, physics-informed machine learning has shown to be a potential solution to this, by embedding known constraints from physics \cite{harder_hard-constrained_2024,harder_physics-informed_2022}.

\textbf{Generative models}, which aim to learn underlying probabilities distributions of the data, are a promising approach for matching downscaled statistics to that of observations (similar to statistical downscaling with ``model output statistics'' described above). Many high resolution images can be generated through sampling, or \textit{conditioning on}, the low resolution data (also making them a type of ``stochastic weather generator''). For example, several recent studies use Generative Adverserial Networks (GANs) to generate high resolution images conditioned on low resolution images  \cite{leinonen_stochastic_2021,harris_generative_2022,oyama_deep_2023,wang_fast_2021,price_increasing_2022}. 
Diffusion-based approaches have become established as a state-of-the-art technique for image generation (e.g., Stable Diffusion \cite{rombach_high-resolution_2022}, DALL-E2 \cite{ramesh_hierarchical_2022}, amongst others \cite{saharia_photorealistic_2022,nichol_glide_2022}) but have not yet become widely used for climate downscaling. We expect to see more diffusion-based downscaling methods in the near future, following \cite{bischoff_unpaired_2023,wan_debias_2023} who used diffusion for downscaling turbulent fluid data and \cite{mardani_generative_2023,addison_machine_2022,nath_forecasting_2024} who presented diffusion for downscaling climate model output on localised scales.
In a similar realm, \cite{Groenke_ClimAlign_2020} use unsupervised normalising flows to downscale climate variables. In this study, we demonstrate the performance of a diffusion-based approach to downscaling on a continental scale.

The probabilistic nature of generative machine learning makes them particularly desirable for risk assessment studies, such as those aiming to quantify the likelihood of extreme events. Both weather and climate modelling communities have long used ensembles of simulations to assess model, scenario, and initial condition uncertainty \cite{hawkins_potential_2009}. There is a growing interest in generative machine learning for generating ensembles from one climate/weather model simulation \cite{li_seeds_2023}. In a downscaling setting, ensembles could be leveraged to determine the trustworthiness of predictions, for example, by highlighting increased model uncertainty when applied to out-of-distribution samples that are likely to occur in a non-stationary climate \cite{fowler_linking_2007}.

\section{Methods}
\label{sec:methods}

\subsection{Data}
\label{subsec:methods-data}

\textbf{The ERA5 reanalysis dataset}, made publicly available by ECMWF \cite{hersbach_era5_2020}, is used in this study. Reanalysis datasets optimally combine observations and models through data assimilation techniques. ERA5 includes hourly estimates for a wide range of atmospheric, land, and oceanic climate variables on a $0.25\degree$~resolution longitude-latitude grid with 137 vertical levels, from January 1940 to present day. This dataset is becoming widely used in other machine learning weather/climate prediction studies \cite{pathak_fourcastnet_2022,bi_pangu-weather_2022,lam_learning_2023,chen_fuxi_2023}.

We consider downscaling of three variables over the continental USA: 
\begin{itemize}
\item Air temperature at 2 m 
\item Zonal wind at 100 m
\item Meridional wind at 100 m
\end{itemize}
We focus only on these three variables over the USA as a demonstration of diffusion for downscaling on continental scales. Although relevant for extreme events, we do not include precipitation here because because ERA5 reanalysis at $0.25\degree$ resolution is likely too coarse to capture its spatial intermittency \cite{bihlo_generative_2021}. Diffusion-based approaches to downscaling precipitation on kilometer scales can be found in \cite{mardani_generative_2023,harris_generative_2022,nath_forecasting_2024}.

ERA5 provides hourly data from 1940 to present day, however, we expect this to be highly correlated in the temporal domain. To reduce the dataset size without significant loss of information, we subsample randomly in time to select only 30 timesteps per month. This reduces the dataset size by $1/24$ and gives us data sampled approximately once per day, at different times of day. We use years 1950-2022 and separate this into a training dataset (1950-2017) and testing dataset (2018-2022). 

We aim to downscale a coarsened version of the ERA5 dataset back onto the original  $0.25\degree$~resolution grid. For the coarsened dataset, we use bi-linear interpolation onto a $2\degree$~resolution grid, to be approximately consistent with typical climate model resolution (e.g., $\mathcal{O}(1\degree)$ in CMIP6 models \cite{chen_framing_2021}). This corresponds to approximately $8\times$ scaling of resolution. Note that this approach assumes the coarse resolution data is not biased, following the perfect prognosis approach to downscaling. To apply this to real climate model data, one may first need to carry out bias correction \cite{mardani_generative_2023}.

\subsection{Baseline U-Net for downscaling}
\label{subsec:methods-u-net}

Our problem is a supervised learning problem, where we aim to learn the fine resolution variables from the coarse resolution variables. We will compare two approaches for this task: firstly, a U-Net architecture and secondly, a diffusion-based generative model. Since the inputs and outputs are both images with a channel for each variable, this requires an image-to-image model such as a U-Net \cite{navab_u-net_2015}. A U-Net architecture is a neural network comprised of several encoding convolutional layers followed by several decoding up-convolutional layers, making them ideal for computer vision tasks when the input and output images are of the same size. They have shown excellent performance on image processing tasks, such as segmentation \cite{navab_u-net_2015} and and super-resolution \cite{hu_runet_2019}. As we will see in the following section, our diffusion-based approach also uses a U-Net, allowing us to use the same base model, making for a fair comparison between these methods. 

Our input variables are the air temperature at 2~m, and zonal and meridional winds at 100~m, all coarsened onto the $2\degree$~resolution grid. Note that after coarsening, these variables are stored on the fine resolution ($0.25\degree$) grid using bi-linear interpolation, ensuring that the inputs and outputs have the same dimensions and allowing for a simple U-Net architecture. To improve efficiency of the U-Net, rather than directly learning the output on the fine resolution grid, we learn the difference between the output on the fine and coarse grid. As well as these inputs, we also provide the U-Net with fixed inputs for spatial quantities that describe the geography and are constant in time. These are the land-sea mask and the height of the land surface, defined on the fine resolution grid. We expect these to aid learning of details around the coast, the great lakes and the mountains. This means the input image is of size (128 x 256 x 5) while the output image is of size (128 x 256 x 3). Finally, we also provide scalar values representing the month and the time of day, allowing the U-Net to learn differences in the diurnal and seasonal cycles. These are input to the each block of the U-Net, using a shallow neural network. Table~\ref{tab:inputs-outputs} shows these inputs and outputs to the network. All variables are normalised to have standard scaling with zero mean and unit variance. The U-Net is trained by minimising the Mean Squared Error (MSE) between the U-Net predicted image and the samples from the data.

\begin{table}
 \caption{Table describing inputs and outputs in this study. By "fixed" inputs we mean the same for all inputs.}
  \centering
  \begin{tabular}{ll}
    \toprule
    Inputs                             & Outputs   \\
    \midrule
    \textbf{Inputs (coarse grid):}     & \textbf{Difference (fine grid - coarse grid):}              \\
    \midrule
    Temperature at 2m level            & Temperature at 2m level     \\
    Zonal winds at 100m level          & Zonal winds at 100m level      \\
    Meridional winds at 100m level     & Meridional winds at 100m level \\
    
    \cmidrule(r){1-1}
    \textbf{Fixed inputs (fine grid):}          & \\
    \cmidrule(r){1-1}
    Land-sea mask                      & \\
    Height of land surface             & \\
    
    \cmidrule(r){1-1}
    \textbf{Embeddings (scalars):}       & \\
    \cmidrule(r){1-1}
    Month                             & \\
    Time of day                       & \\
    
    \bottomrule
  \end{tabular}
  \label{tab:inputs-outputs}
\end{table}

\subsection{Diffusion-based generative model}
\label{subsec:methods-ddpm}
Generative models using diffusion have shown great success as a method for synthesising images \cite{sohl-dickstein_deep_2015}.
As a generative model, the goal is to learn some probability distribution $p(\mathbf{x})$ given a set of samples $\{\mathbf{x}\}$. This would be an unconditional model, however, we are interested in building a conditional model, $p(\mathbf{x} | \mathbf{y})$, where $\mathbf{x}$ are the high resolution image samples and $\mathbf{y}$ are the low resolution image samples.
In the following discussion, we will neglect the conditioning on $\mathbf{y}$ for clarity, but note that the conditional form is obtained by replacing $\mathbf{x}$ with $\mathbf{x} | \mathbf{y} $.

Diffusion models take inspiration from thermodynamics whereby a diffusion process $\{\mathbf{x}(t)\}^T_{t=0}$ with $t \in [0, T]$ is constructed which transforms samples from the data distribution $\mathbf{x}(0) \sim p(\mathbf{x})$ to that of an isotropic Gaussian $\mathbf{x} (T) \sim \mathcal{N}(\mathbf{0}, \sigma \mathbf{I})$.
Following \cite{song_score-based_2021}, the stochastic differential equation for such a process is given by
\begin{equation}
    \mathrm{d} \mathbf{x} = \mathbf{f} (\mathbf{x}, t) \mathrm{d} t + g(t) \mathrm{d} \mathbf{w}
\end{equation}
where $\mathbf{f} (\mathbf{x}, t)$ is the drift coefficient $g(t)$ is the diffusion coefficient and $\mathbf{w}$ is a Wiener process.
If this process can be reversed, one can use it to generate samples from $p(\mathbf{x}_0)$.
This involves first sampling from the terminal distribution $\mathbf{x}(T)$ (the isotropic Gaussian), then transforming the sample through the reverse processes to the data distribution.
It can be shown that the reverse of a diffusion process is itself a diffusion process with the following reverse time stochastic differential equation (SDE) \cite{anderson_reverse-time_1982} 

\begin{equation}
        \mathrm{d} \mathbf{x} = [\mathbf{f} (\mathbf{x}, t) - g(t)^2 \nabla_\mathbf{x} \, \mathrm{log} \, p_t(\mathbf{x})] \mathrm{d} t + g(t) \mathrm{d} \overline{\mathbf{w}}
\end{equation}
where $\overline{\mathbf{w}}$ is a Wiener process with time flowing backwards from T to 0. 
$\nabla_\mathbf{x} \mathrm{log} p_t(\mathbf{x})$ is the score function, a vector field which points towards higher density. 
The score does not depend on the intractable normalisation constant making it easier to evaluate.
Obtaining samples by solving the reverse SDE requires knowledge of the score function.
To do this we parameterise the score using a neural network $\mathbf{s}_\mathbf{\theta} (\mathbf{x}, t)$, with the same U-Net architecture as described above in Section \ref{subsec:methods-u-net}.
The weights of $\mathbf{s}_\mathbf{\theta}(\mathbf{x}, t)$ are optimised using score matching \cite{vincent_connection_2011} by minimising the following loss \cite{song_score-based_2021} 
\begin{equation}
    \boldsymbol{\theta}^* = \underset{\boldsymbol{\theta}}{\mathrm{arg \, min}} \,\,\mathbb{E}_t\Big[ \lambda(t) \mathbb{E}_{\mathbf{x}(0)} \mathbb{E}_{\mathbf{x}(t) | \mathbf{x}(0)} \Big[\| \mathbf{s}_\mathbf{\theta} (\mathbf{x}, t) - \nabla_{\mathbf{x}(t)} \mathrm{log} \, p_{t}(\mathbf{x}(t) | \mathbf{x}(0) )\|^2_2 \Big] \Big]
\end{equation}
where $\lambda(t)$ is a weighting function, t is sampled uniformly between $[0, T]$. This equation is obtained through minimising the evidence lower bound (ELBO) on the negative log likelihood, $\mathbb{E} [-\mathrm{log}\,p(\mathbf{x}_0)]$, and reweighting by $\lambda(t)$.

Within the framework discussed above, there are a number of design choices and free parameters to set.
In this work, we use the implementation given by Karras et al \cite{karras_elucidating_2022}.
Their suggestions give a noticeable improvement over previous works  \cite{song_score-based_2021,ho_denoising_2020,nichol_improved_2021} for image synthesis.
One improvement to note is the use of a higher order (second)  integration scheme.
This significantly reduces the number of timesteps taken when solving the reverse process making inference significantly more efficient. We use 100 timesteps during evaluation, but find that using just 50 is sufficient for a similar accuracy.

\section{Results}
\label{sec:results}

After training on years 2050-2017 of ERA5, we evaluate the performance of both the U-Net and the diffusion-based model on years 2018-2022.

\subsection{Example}
\label{subsec:example-plots}

\begin{figure}
  \centering
  \includegraphics[width=\textwidth]{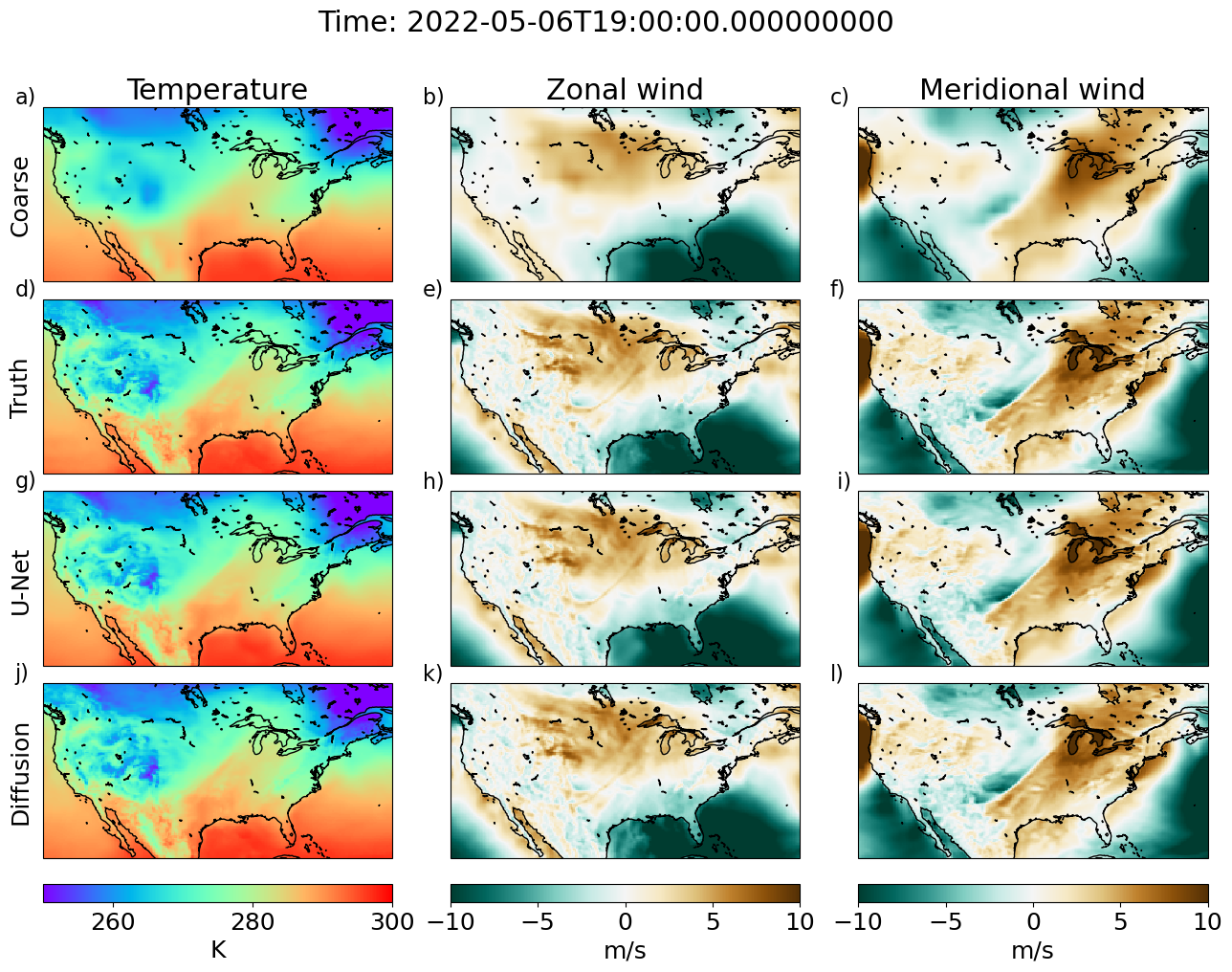}
  \includegraphics[width=\textwidth]{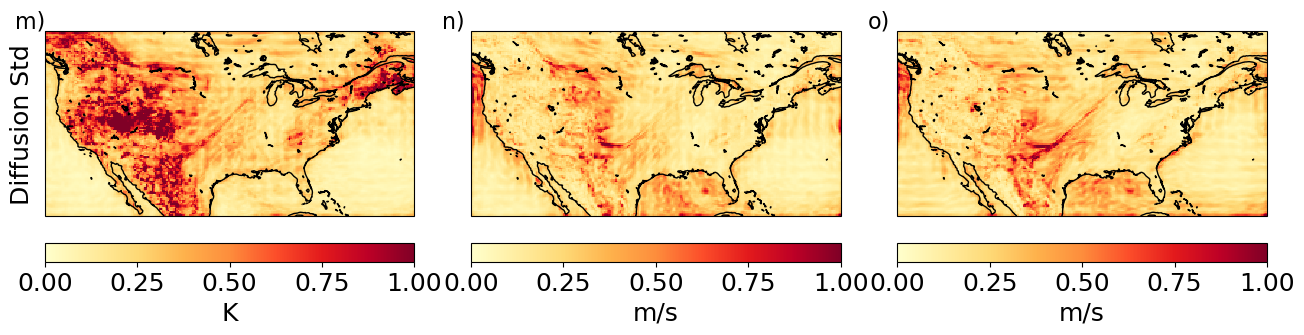}
  \caption{Maps comparing (a-c) coarse resolution (the input) to the fine resolution for (d-f) the truth, (g-i) the U-Net downscaled prediction and (j-l) one generated diffusion downscaled prediction, for all three variables at a single timestep. (m-n) show 1 standard deviation across a 30-member ensemble generated by diffusion.}
  
  \label{fig:example-timestep}
  
\end{figure}

Figure \ref{fig:example-timestep} shows the coarse resolution maps, the true fine resolution maps, and the downscaled fine resolution maps for the U-Net and diffusion, for one snapshot for all three variables. Both machine learning approaches demonstrate good predictions when compared against the fine resolution. For temperature, the U-Net and diffusion predictions are virtually indistinguishable. For the winds, there are some smaller scale features in the diffusion prediction, while the U-Net predicts smoother features. We expect this is because U-Net is trained to minimise MSE, which naturally favours smoother fields rather than high frequency variations.

\subsection{Probabilistic Predictions}
\label{subsec:probalistic}

Figure \ref{fig:example-timestep}(j-l), shows one generated downscaled prediction from the diffusion model. However, one of the benefits of the diffusion-based model lies in its ability to generate multiple predictions, creating probability distributions, rather than one deterministic prediction. We generate an ensemble of diffusion predictions consisting of 30 samples, each of which is generated from a different sample from the terminal Gaussian distribution $\mathbf{x}(T)$. This provides us with valuable information when making high-stakes decisions, for example, by highlighting when there is higher uncertainty amongst ensemble members. We find the ensemble members agree reasonably well and show their 1 standard deviation in Figure \ref{fig:example-timestep}(m-n) for each variable. There is higher uncertainty in regions where the variable is changing quickly, in other words, where there are larger spatial gradients. For the surface temperatures, this increased uncertainty tends to be over mountainous regions and for the winds, to the east of mountains and along fronts where the winds are rapidly changing. This makes sense for a downscaling problem where the prediction is constrained by the coarse grid, but the diffusion generates many possible realisations for how the variables could be interpolated between the coarse grid cells. This also results in a gridded pattern where predictions closer to the coarse grid cells have lower uncertainty.

\subsection{Metrics}
\label{subsec:metrics}

Here, we aim to compare the results of a deterministic U-Net with a probabilistic diffusion-based model. To compare both, we will use the mean absolute error (MAE). For the diffusion-based model, we use the mean across across the ensemble. To consider the probabilistic nature of diffusion, we will also consider the continuous ranked probability score (CRPS), a generalisation of mean absolute error to compare predicted probability distributions to a single ground truth \cite{hersbach_decomposition_2000,gneiting_strictly_2007}. Note, that we present the CRPS metric for diffusion only and cannot fairly compare this with the deterministic U-Net predictions. 

\begin{figure}
  \centering
  \includegraphics[width=\textwidth]{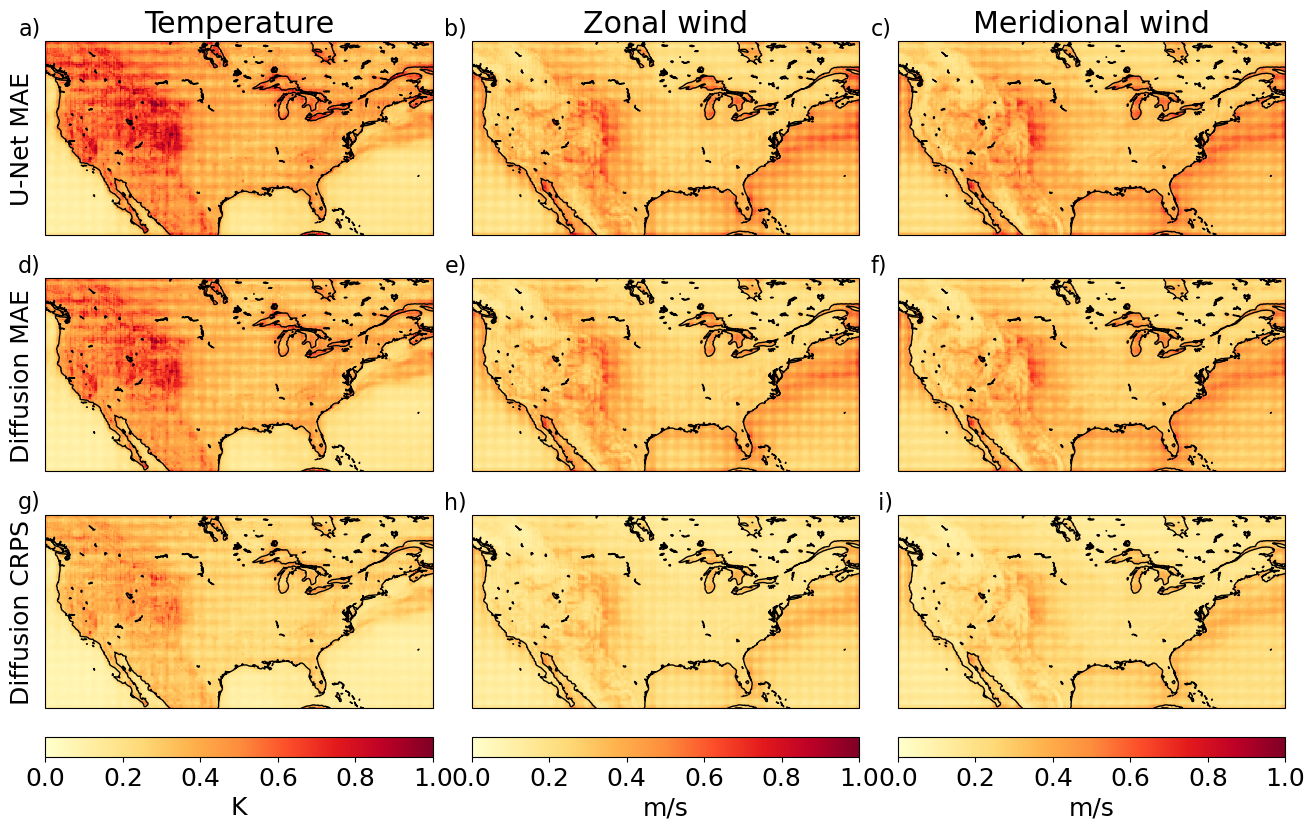}
  \caption{Maps comparing the error metrics for the U-Net downscaled prediction and the diffusion downscaled prediction, for all three variables. }
  \label{fig:error-metric-maps}
\end{figure}

Figure \ref{fig:error-metric-maps} shows maps showing these metrics averaged over the entire test dataset. The gridded pattern appears due to reduced error at grid-cells close to the coarse-reoslution grid, also present in Figure \ref{fig:example-timestep}(m-o).  The diffusion-based approach shows slightly lower MAE in the high altitude regions along the Rockies, particularly for temperature. These are the regions that exhibit more variability, both in time and space, which could explain why a probablistic model performs better. The value in the probabilistic approach is further highlighted by the significant reduction in error when considering the CRPS metric for diffusion, which takes into account all ensemble members.

Table \ref{tab:error-metrics} shows the mean metrics, averaged across the entire domain and test dataset. Note that when taking spatial averages, we weight each grid-cell by the cosine of latitude to account for the non-uniform size of grid cells.

\begin{table}
 \caption{Results}
  \centering
  \begin{tabular}{lllll}
    \toprule
    Metric & Model                             & Temperature ($\degree$C)                &  Zonal winds  (m/s) & Meridional winds (m/s) \\
    \midrule
    MAE & Linear Interpolation     & 1.047 & 0.733 & 0.746 \\
    & U-Net                    & 0.384 & 0.335 & 0.348 \\
    & Diffusion                & \textbf{0.328} & \textbf{0.308} & \textbf{0.319} \\
    \midrule
    CRPS & Diffusion               & 0.254 & 0.224 & 0.232 \\
    \bottomrule
  \end{tabular}
  \label{tab:error-metrics}
\end{table}

\subsection{Spectra}
\label{subsec:spectra}

In the example snapshots (Figure \ref{fig:example-timestep}), it appears that the diffusion model produces more accurate high frequency variations for the winds compared to the U-Net. We validate this further by calculating the power spectra across all wavelengths, for all three variables. 
Here, we treat the data as an image and take a 2D Fourier transform across this image to estimate the power for each wavelength. These results are robust to transforming the data onto a grid equispaced in distance, accounting for different grid spacing in latitude.

\begin{figure}
  \centering
  \includegraphics[width=\textwidth]{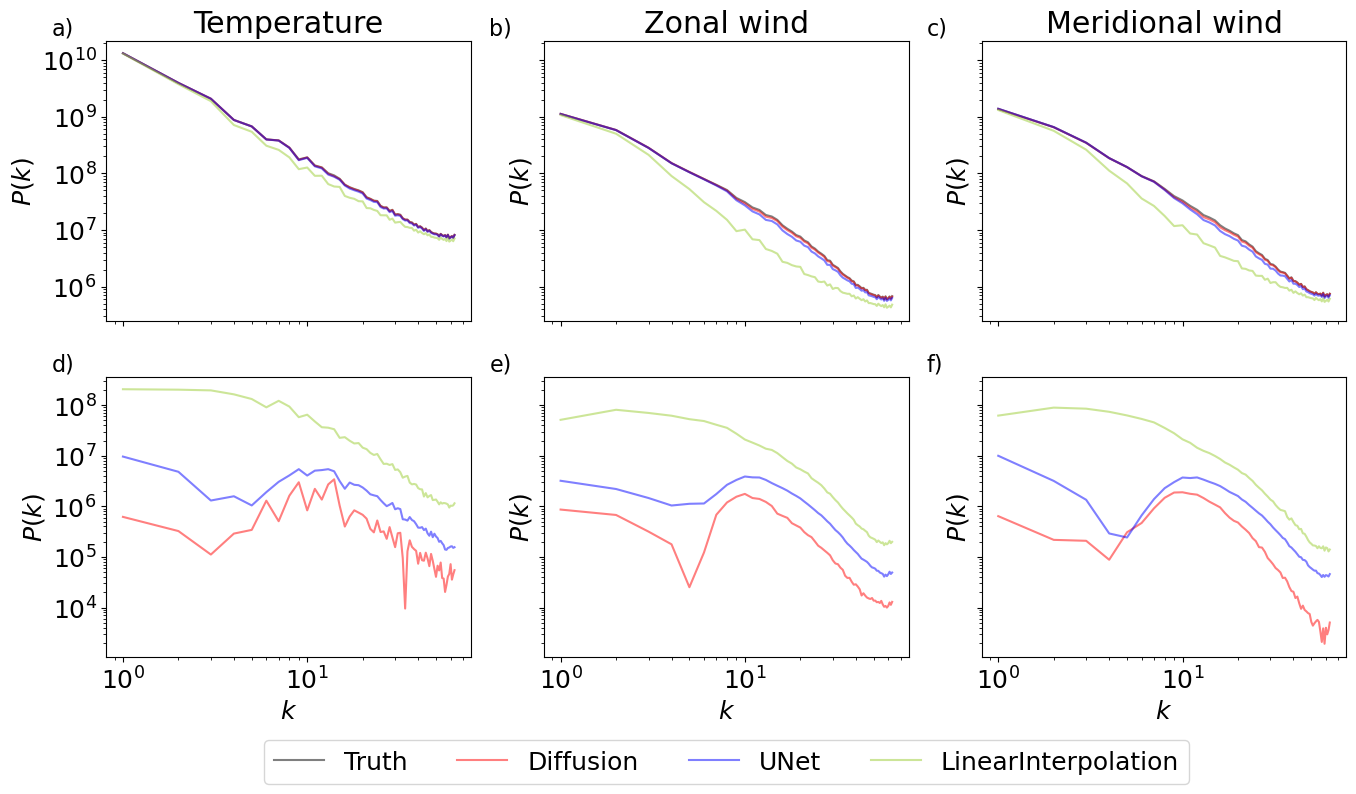}
  \caption{(a-c) Power spectrum on a log-scale for all three variables for the truth, diffusion, U-Net and linear interpolation, and (d-f) the difference between the truth and the predicted power spectra for diffusion, U-Net and linear interpolation.}
  \label{fig:spectra}
\end{figure}

Figure \ref{fig:spectra}(a-c) shows the power spectra for each variable for all methods on a log-scale, compared against the ground truth in black. The power spectra for diffusion is indistinguishable from the ground truth, so we also present the differences between these in Figure \ref{fig:spectra}(d-f). The power spectra is significantly more accurate across all scales in the diffusion model in comparison to the U-Net. For the winds, we see this is more evident at the high wavenumbers. This shows the promise of diffusion for approximating geospatial data for a range of uses, potentially addressing issues of oversmoothing seen in other weather and climate studies \cite{pathak_fourcastnet_2022,bi_pangu-weather_2022,lam_learning_2023}.

\section{Conclusions}
\label{sec:conclusions}

In this paper, we have presented a generative diffusion-based model for downscaling climate data on continental scales. We demonstrated this by recovering $0.25\degree$~resolution ERA5 data from a $2\degree$~resolution coarse-grained version of ERA5 and found that the diffusion model outperformed a baseline U-Net with the same architecture. The next step would be to apply this diffusion model to downscale the output of a coarse $\mathcal{O}($1\degree$)$~resolution climate model to a higher resolution, e.g., $0.25\degree$~ERA5. This presents the additional challenge that the coarse model output may not match up with the high resolution dataset. One approach could be to first apply a bias-correction technique, as done in \cite{mardani_generative_2023}. The application of downscaling to climate model output based on historical observational datasets also presents the issue of non-stationarity, whereby we cannot assume that the relationship between the coarse resolution and the high resolution data remains constant under a changing climate. Diffusion-based approaches may be appealing for this task, as they predict an ensemble which can highlight the epistemic uncertainty associated with each prediction.

There are a wide range of avenues to expand upon this study, for example, increasing the number of variables predicted, exploring performance at different resolutions, applying the same method to different regions or to full global data, or incorporating temporal, as well as spatial, downscaling. Downscaling of precipitation is of particular interest, due to its intermittent spatial patterns that is typically overly smooth in climate model output compared to observations and is a crucial variable for extreme events such as flooding \cite{sunyer_comparison_2012,bano-medina_downscaling_2022,vandal_deepsd_2017,xu_precipatch_2020,akinsanola_evaluation_2018,babaousmail_novel_2021,harris_generative_2022}.

This study was carried out without access to expensive, high performance computing systems. Although both training and inference was more expensive than the baseline U-Net, diffusion offers significantly improved performance and the ability to generate ensembles, with a computational cost several orders of magnitude lower than full climate model simulations. This could make impact studies accessible and affordable to a wider range of end-users. We conclude this study by noting that, given the rapid improvements in diffusion-based image generation in the last three years \cite{zhang_text--image_2023}, we might expect to see even more advances in diffusion-based techniques applied to climate problems. Even during the time it took us to complete this study, we have seen diffusion models gaining popularity in this field \cite{mardani_generative_2023,price_gencast_2023,nath_forecasting_2024,huang_diffda_2024,chan_hyper-diffusion_2024}, a trend that we hope will continue in the years to come.

\section*{Open data statement}
All code used in this study is available at \url{https://github.com/robbiewatt1/ClimateDiffuse}. We followed the diffusion implementation of \cite{karras_elucidating_2022} available at \url{https://github.com/NVlabs/edm}.
The ERA5 reanalysis dataset used in this study is publicly available at \url{https://cds.climate.copernicus.eu/cdsapp#!/dataset/reanalysis-era5-complete?tab=overview} \cite{hersbach_era5_2020}.

\section*{Acknowledgements}
This was done in our spare time and not directly funded. We acknowledge our employer, Stanford University, for support and access to journal articles. LAM would also like to acknowledge support from Schmidt Sciences, that funds her research at the intersection of machine learning and climate modeling.

\bibliographystyle{apalike}  
\bibliography{references}

\end{document}